\documentclass[10pt,letterpaper,twocolumn]{article}
\usepackage[latin9]{inputenc}
\usepackage{amsmath}
\usepackage{amssymb}
\usepackage{graphicx}
\usepackage[unicode=true,pdfusetitle,
 bookmarks=true,bookmarksnumbered=false,bookmarksopen=false,
 breaklinks=false,pdfborder={0 0 1},backref=section,colorlinks=false]
 {hyperref}
\hypersetup{
 draft}
\usepackage{breakurl}

\makeatletter

\usepackage{ol2}

\makeatother

\begin{document}
\twocolumn[
\title{Observation of spectral interference for any path difference in an interferometer}
\author{Luis Jos\'{e} Salazar-Serrano,$^{1,2}$ Alejandra Valencia$^2$ and J.P. Torres$^{1,3}$}
\address{
$^1$ICFO-Institut de Ciencies Fotoniques, Mediterranean Technology
Park, 08860 Castelldefels (Barcelona), Spain \\
$^2$ Quantum Optics Laboratory, Universidad de los Andes, AA 4976,
Bogot\'{a}, Colombia \\
$^3$ Department of Signal Theory and Communications, Universitat
Politecnica de Catalunya, 08034 Barcelona, Spain

$^*$Corresponding author: luis-jose.salazar@icfo.es}

\begin{abstract}
We report the experimental observation of spectral interference in
a Michelson interferometer, regardless of the relationship between
the temporal path difference  introduced between the arms of the
interferometer and the spectral width of the input pulse. This
observation is  possible by introducing the polarization degree of
freedom into a Michelson interferometer using a typical weak value
amplification scenario.
\end{abstract}


\ocis{260.0260, 260.3160,260.5430, 320.5550,120.3180}

]


\noindent 

Interference is a fundamental concept in any theory based on
waves, such as classical electromagnetism or quantum theory. The
specific experimental arrangement required for the observation of
interference depends on the characteristics of the light source,
i.e., its spatio-temporal profile and its degree of coherence. For
example, for first-order coherent light in a Michelson
interferometer, for temporal delays shorter than the pulse width,
interference manifests as a delay-dependent change of the
intensity at the output port of the interferometer. For longer
temporal delays, interference manifest as spectral interference
for a given temporal delay. The observation of spectral
interference was denoted by Mandel ~\cite{givens1961,mandel1962}
as the Alford-Gold effect~\cite{alford_gold1958} and it is well-
known in optics~\cite{zou_mandel1992}.

Here we report the observation of spectral interference
independently of the temporal regime under consideration. The
interference is revealed as a reshaping of the input spectrum that
is accomplished by introducing the polarization degree of freedom
into a Michelson interferometer. This scenario corresponds
precisely to the conditions of a typical weak value amplification
configuration~\cite{aharonov1988,duck1989,boyd2013,jordan2013}
that although was originally conceived in the framework of a
quantum formalism, it is essentially based on the phenomena of
interference and can thus be applied to any scenario with waves
\cite{howell2010, brunner2010, li_guo2011, xu_guo2013,
salazar2014}.

\begin{figure}[t!]
\centering
\includegraphics[width=0.45\textwidth]{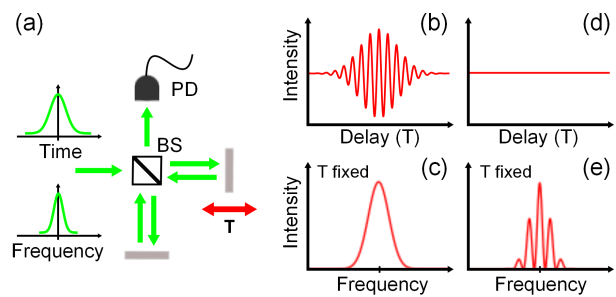}
\caption{Typical interference patterns that can be observed in a
Michelson interferometer, in the time and frequency domains. (a)
depicts the interferometric configuration considered. BS: beam
splitter, PD: Photodetector. (b) and (d) show the intensity
measured as a function of a temporal delay $T$. (c) and (e) show
the intensity as a function of frequency for a given value of $T$.
(b) and (c) correspond to the case $T \ll \tau$, while (d) and (e)
correspond to the case, $T \gg \tau$.} \label{michelson-setup}
\end{figure}

For the sake of clarity, let us start by describing temporal and
spectral interference in a typical Michelson interferometer,
without considering polarization. Later on, we will describe the
effects that the introduction of the polarization degree of
freedom has on spectral interference.  Consider the situation
depicted in Fig.~\ref{michelson-setup}(a). A first-order coherent
input pulse with amplitude $E_0$, central frequency $\nu_0$, input
polarization ${\bf e}_{in}$, and temporal duration $\tau$
(full width at half maximum) described by
\begin{equation}
{\bf E}_{in}(t) = E_0\,\exp \left[-(2\ln
2/\tau^2)\,t^2+i\,2\pi\nu_0 t\right]\,{\bf e}_{in}\,, \label{eq:EinitialTime}
\end{equation}
enters a Michelson interferometer where is divided in two beams by
a beam splitter (BS). Each of the new pulses follows a different
path and after reflection in a mirror the two pulses recombine at
the BS. By changing the position of one of the mirrors, a temporal
delay ($T$) is generated between the two pulses. The intensity
measured by a slow detector at the output port of the
interferometer as a function of $T$ can be written as
\begin{equation}
I(T) = \frac{I_0}{2}\,\left[ 1 + \exp \left(-\ln2\,T^2/\tau^2
\right)\,\cos\left(2\pi\nu_0T\right)\right]\,,
\label{intensity-time-general}
\end{equation}
where $I_0 = |E_0|^2$. Two interesting cases can be distinguished:
i) when $T \ll \tau$, and therefore, the two pulses traveling the
different paths overlap in time and ii) when $T \gg \tau$ and the
two pulses do not overlap. In the first case, the output intensity
as a function of $T$ reduces to
\begin{equation}\label{temporal interference Tsmall}
I_{ T \ll \tau}(T) = \frac{I_0}{2}\,[1+\cos(2\pi\nu_0T)],
\end{equation}
while in the case $T \gg \tau$, Eq.~(\ref{intensity-time-general})
becomes
\begin{equation}\label{temporal interference Tbig}
I_{T \gg \tau}(T) = \frac{I_0}{2}\,.
\end{equation}
These results are very well known in optics. They indicate the
presence of temporal interference for the case $T \ll \tau$ and
its absence for $T \gg \tau$. The two situations are illustrated
in Fig.~\ref{michelson-setup}(b) and
Fig.~\ref{michelson-setup}(d).

An analogous analysis can be done when the detector in
Fig.~\ref{michelson-setup}(a) is changed by a spectrometer and the
power spectrum $S(\nu)$ as a function of the frequency $\nu$ is
considered. In this case,
\begin{equation}\label{spectral-interference-no-polarization}
S(\nu) = \frac{S_{in}(\nu)}{2}\left[1+\cos\left(2\pi\nu\,T \right)\right]\,,
\end{equation}
where
\begin{equation}\label{frequency interference Tsmall}
S_{in}(\nu)=S_0\exp\left[-(\pi^2\tau^2/\ln2)
(\nu-\nu_0)^2\right]\,
\end{equation}
is the input power spectrum with $S_0$ being a constant.

Equation~(\ref{spectral-interference-no-polarization}) indicates
that $S(\nu)$ corresponds to a reshaping of the input spectrum.
This reshaping can be understood if $h(\nu)=\tfrac{1}{2}[1+\cos(2\pi\nu\,T)]$ is
considered as a transfer function that describes the effect of the
interferometer. Therefore depending on the relationship between
the oscillation frequency of the transfer function, defined by
$T$, and the pulse width, $\tau$, it is possible to obtain
different reshapings of the input spectrum that can be identified
with interference. Fig.~\ref{michelson-setup}(c) and
Fig.~\ref{michelson-setup}(e) depict the output power spectrum for
the regime $T \ll \tau$ and $T \gg \tau$, respectively. The two
cases are clearly different. As expected from standard
interferometry, when $T \ll \tau $,  the output power spectrum, is
identical to the input one, while for $T \gg \tau$, a clear
reshaping of the input spectrum appears. This last case
corresponds to the Alford-Gold effect and indicates that a
Michelson interferometer can be seen as a periodic filter that
produces a modulation of the initial spectrum with a peak
separation proportional to $\sim 1/T$.

\begin{figure}[t!]
\centering
\includegraphics[width=0.45\textwidth]{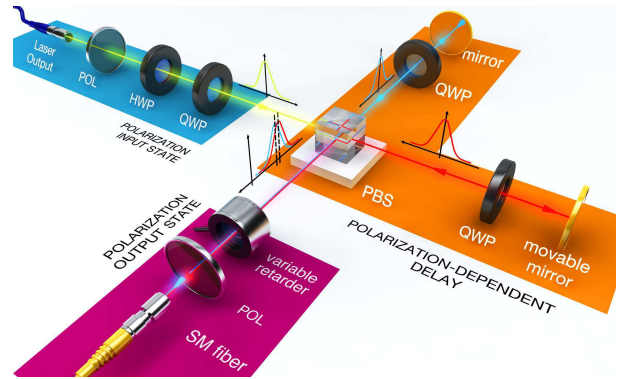}
\caption{Experimental setup. The input pulse polarization state is
selected to be left-circular by using a polarizer, a quarter wave
plate (QWP) and a half wave plate (HWP). A Polarizing beam
splitter (PBS) splits the input into two orthogonal linear
polarizations that propagate along different arms of the
interferometer. An additional QWP is introduced in each arm to
rotate the beam polarization by $90^{\circ}$ to allow the
recombination of both beams in a single beam by the same PBS. The
output beam spectrum is measured by an optical spectrum analyzer
(OSA) connected to a single mode fiber (SM).}
\label{michelson-weak-setup}
\end{figure}

Now let us describe the effects of adding the polarization degree
of freedom to the Michelson interferometer. For this, consider the
situation depicted in Fig.~\ref{michelson-weak-setup} in which
four main differences with the standard Michelson interferometer
of Fig.~\ref{michelson-setup}(a) can be observed: $1)$ the
presence of polarization control elements in the input port of the
interferometer, $2)$ the substitution of the beam splitter (BS) by
a polarizing beam splitter (PBS), $3)$ the introduction of a
quarter wave plate (QWP) in each arm of the interferometer, and
$4)$ the presence, in the output port, of a variable polarization
analyzer, composed by a liquid crystal variable retarder (LCVR)
and a polarizer at $45^{\circ}$. The polarizer, half-wave plate
and quarter wave plate in the interferometer's input port are used
to set up the polarization of the input beam. The PBS divides
spatially the two orthogonal polarization components of the input
beam and the QWP rotates the corresponding polarization component
by $90^{\circ}$ after reflection in each mirror. The movable
mirror generates the temporal delay $T$.

Since two beams with orthogonal polarizations do not interfere, an
interference effect is generated by the polarization control
elements of the output port that generate an additional phase
difference, $\Gamma$, between the pulses coming from the two paths
of the interferometer. In particular, for a left-handed circularly
polarized input beam $({\bf e}_{in}={\bf x}-i {\bf y})$, where
${\bf x}$ denotes horizontal polarization and ${\bf y}$ vertical
polarization, the output electric field reads
\begin{eqnarray}
& & {\bf E}_{out}(\nu) = E_0\sqrt{\frac{\pi\tau^2}{4\ln2}}\exp\left\{-[\pi^2\tau^2/(2\ln2)]\,(\nu-\nu_0)^2\right\} \nonumber \\
& & \times \left\{ \exp\left( -i\,2\pi\nu T \right){\bf
x}+\exp\left[ -i\,\left(\Gamma+\pi/2\right)\right]{\bf
y}\,\right\}.
\end{eqnarray}
The power spectrum of the light at the interferometer's output
port is then given by
\begin{equation}
S_{out}(\nu)
=\frac{S_{in}(\nu)}{2}\left[1+\cos\left(2\pi\nu\,T-\Gamma-\pi/2\right)\right]\,.
\label{eq:SpectrumOutTime}
\end{equation}

\begin{figure}[t!]
\centering
\includegraphics[width=0.45\textwidth]{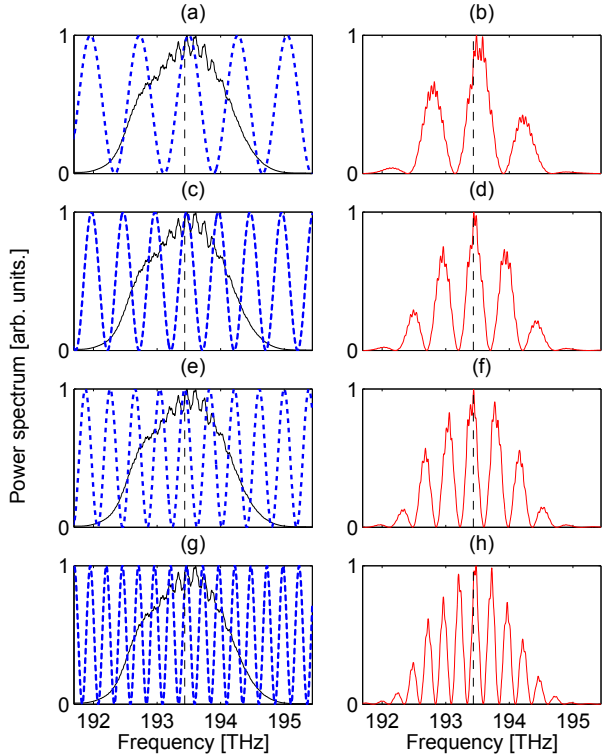}
\caption{Spectral interference in the regime $T\gg\tau$. First
column depicts theoretical transfer function, $H(\nu)$, as a
dashed line for $\Gamma = 61.5^{\circ}$ and different time delays:
$T=1453\,\mathrm{fs}$ in (a), $T=2120\,\mathrm{fs}$ in (c),
$T=2786\,\mathrm{fs}$ in (e) and $T=4120\,\mathrm{fs}$ in (g). The
solid line corresponds to the measured spectrum of a pulsed laser
with a temporal duration $\tau=320\,\mathrm{fs}$ centered in
$\nu_0=193.5\,\mathrm{THz}$. Second column shows the experimental
results for $S_{\mathrm{out}}(\nu)$. The vertical dashed line
indicates the central frequency of the input spectrum.}
\label{experiment-alford-gold}
\end{figure}

\begin{figure}[t!]
\centering
\includegraphics[width=0.45\textwidth]{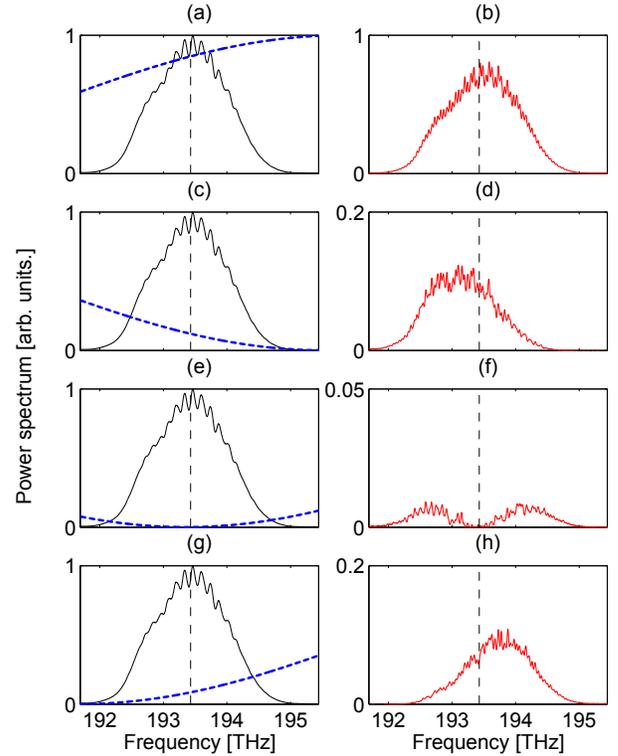}
\caption{Spectral interference in the regime $T\ll\tau$. First
column depicts theoretical transfer function, $H(\nu)$, as a
dashed line for a fixed delay of $T=53.7\,\mathrm{fs}$ and
$\Gamma=98.5^{\circ}$ in (a), $\Gamma = 273.1^{\circ}$ in (c),
$\Gamma = 231.2^{\circ}$ in (e), and $\Gamma = 198.9^{\circ}$ in
(g). The solid line corresponds to the measured spectrum of a
pulsed laser with a temporal duration $\tau=320\,\mathrm{fs}$
centered in $\nu_0=193.5\,\mathrm{THz}$. Second column shows the
experimental results for $S_{\mathrm{out}}(\nu)$. The vertical
dashed line indicates the central frequency of the input spectrum.
To help the eye the vertical axis has been reescaled since the
output intensity changes dramatically with different values of
$\Gamma$ in this regime.} \label{experiment-weak-Tll}
\end{figure}

In the same way as was done for
Eq.~(\ref{spectral-interference-no-polarization}), it is possible
to identify from Eq.~(\ref{eq:SpectrumOutTime}) a transfer
function, $H(\nu)=\tfrac{1}{2}[1+\cos(2\pi\nu\,T -\Gamma-\pi/2)]$,
and distinguish two cases depending on the relationship between
the frequency of oscillation of $H(\nu)$  and the width of the
input power spectrum. We illustrate in the first column of
Figs.~\ref{experiment-alford-gold} and \ref{experiment-weak-Tll}
various transfer functions (dashed lines) for different values of
$\Gamma$ and $T$  and the measured power spectrum of a pulsed
laser (solid line) with a temporal duration
$\tau=320\,\mathrm{fs}$ and central frequency
$\nu_0=193.5\,\mathrm{THz}$. We observe that while for the case $T
\gg \tau$ (Fig.~\ref{experiment-alford-gold}) various oscillations
of the transfer function fit inside the initial spectrum
bandwidth, for $T \ll \tau $ (Fig.~\ref{experiment-weak-Tll}) this
is not the case. This fact results in an output spectrum that
corresponds to different kinds of reshapings of the input spectrum
depending on the regime $T \gg \tau$ or $T \ll \tau$. However, it
is important to notice that for all regimes the reshaping of the
spectrum  is the result of an interference effect and therefore
indicates that spectral interference can be observed in a
Michelson interferometer, independently of the temporal path
difference under consideration, when the polarization degree of
freedom is considered in the interferometer.

To check experimentally that spectral interference can be observed
independently of the value of $T$, the setup of
Fig.~\ref{michelson-weak-setup}, based only on linear optical
elements, was implemented. An input pulse (Calmar Laser -
Mendocino) with a temporal duration $\tau=320\,\mathrm{fs}$,
repetition rate of $20\,\mathrm{MHz}$ and central frequency
$\nu_0=193.5\,\mathrm{THz}$ ($1550\,\mathrm{nm}$) is prepared with
left-handed circular polarization, ${\bf e}_{in}={\bf x}-i\,{\bf
y}$, by using the combination of a polarizer, HWP and QWP. The PBS
splits the input beam into two beams with orthogonal polarizations
and, as mentioned before, each polarization component is rotated
by $90^{\circ}$ by using the QWP and the corresponding a mirror.
The movable mirror, mounted on a translation stage, allows to
change the temporal delay $T$ between pulses with orthogonal
polarizations. The PBS recombines the two reflected beams in a
single beam that emerges from the PBS, passes through the LCVR
followed by a polarizer, and is finally focused into a single mode
fiber (SM) and its spectrum is measured with an Optical Spectrum
Analyzer (Yokogawa - AQ6370).

The second column of Figs.~\ref{experiment-alford-gold} and
\ref{experiment-weak-Tll} shows the experimental results for $T
\gg \tau $ and $T \ll \tau $, respectively. A reshaping of the
spectrum is clearly observed in both cases. For the case of $T \gg
\tau $, a clear modulation with a spacing between peaks
proportional to $1/T$ is observed corresponding to the Alford-Gold
effect. For the regime $T \ll \tau $, two situations can be
distinguished, both accompanied by different amount of losses. In
some cases the reshaping corresponds to a modulation of the
initial spectrum [Fig.~\ref{experiment-weak-Tll}(f)], while in
other the reshaping of the spectrum translate into a measurable
shift of the central frequency of the output power spectrum when
compared with the central frequency of the input pulse
[Fig.~\ref{experiment-weak-Tll}(d) and (h)]
\cite{brunner2010,xu_guo2013,salazar2014}.

In conclusion, we have shown that spectral interference can also
be observed in the regime of small
optical path differences ($T \ll \tau$). This result complements
the observation of the Alford-Gold effect, which reveals
interference in the frequency domain in the opposite regime, $T
\gg \tau$.  The observation of spectral interference, regardless
the temporal path difference, was made possible by introducing the
polarization variable in a Michelson interferometer in an
experimental scheme similar to the ones used in a weak measurement
scenario. The presence of interference in the regime $T\gg \tau$
appears as a clear modulation of the frequency spectrum with
frequency $\sim 1/T$ [see Figs.~\ref{experiment-alford-gold}(b),
(d), (f) and (h)], while in the case $T \ll \tau$, interference
manifests as a modulation of the input spectrum
[Fig.~\ref{experiment-weak-Tll}(f)] or as a shift of the central
frequency [Figs.~\ref{experiment-weak-Tll}(b), (d) and (h)].

\vspace{1cm} \noindent {\bf Acknowledgements}: We acknowledge
support from the Spanish government projects FIS2010-14831 and
Severo Ochoa programs, and from Fundaci\'o Privada Cellex,
Barcelona. LJSS and AV  acknowledges support from Facultad de
Ciencias, Universidad de los Andes Bogot\'{a}, Colombia.

\end{document}